\documentclass[12pt]{article}

\usepackage{fullpage,sak,latexsym,amsmath,amssymb,epsfig}
\newfontobj{\class}{\bf}

\class{NP}
\class{PH}
\class{NQP}
\class{FP}
\class{BPP}
\class{BQP}
\class{EQP}
\class{AWPP}
\class{AAWPP}
\class{GUM}
\class{GAM}
\class{GUAM}
\class{PUM}
\class{PAM}
\class{PUAM}
\class{UP}
\class{AM}
\class{MA}
\class{PP}
\class{NC}
\class{AC}
\class{ACC}
\class{QNC}
\class{EQNC}
\class{NQNC}
\class{BQNC}
\class{QACC}
\class{QAC}
\class{NQAC}
\class{NQACC}
\class{TC}
\class{QTC}
\class{QMA}
\class{SZK}
\newcommand{\cqs}{{\rm CQS}}

\newcommand{\prob}[1]{{\sc #1}}
\newcommand{\OC}{\prob{Orbit Coset}}
\newcommand{\Stab}{\prob{Stabilizer}}
\newcommand{\OS}{\prob{Orbit Superposition}}
\newcommand{\GM}{\prob{Group Membership}}
\newcommand{\OV}{\prob{Order Verification}}
\newcommand{\Gint}{\prob{Group Intersection}}
\newcommand{\GI}{\prob{Graph Isomorphism}}
\newcommand{\HS}{\prob{Hidden Subgroup}}

\newcommand{\DCM}{\prob{Double Coset Membership}}
\newcommand{\DCNM}{\prob{Double Coset Nonmembership}}
\newcommand{\GF}{\prob{Group Factorization}}
\newcommand{\CI}{\prob{Coset Intersection}}
\newcommand{\CQS}{\prob{Circuit Quantum Sampling}}

\newcommand{\isoto}{\cong}

\newcommand{\co}[1]{{\rm co}{#1}}

\newcommand{\map}[3]{{#1}:{#2}\rightarrow{#3}}

\newcommand{\ket}[1]{|{#1}\rangle}
\newcommand{\tuple}[1]{\langle{#1}\rangle}

\newtheorem{definition}{Definition}[section]

\newtheorem{theorem}[definition]{Theorem}
\newtheorem{corollary}[definition]{Corollary}

\newenvironment{proof}{\noindent{\bf Proof.}}{\hfill$\Box$ \bigskip}

\title{Quantum Algorithms for a set of Group Theoretic Problems}

\author{Stephen A. Fenner\thanks{Computer Science and Engineering
Department, Columbia, SC 29208 USA.  Email {\tt
$\{$fenner$|$zhang29$\}$@cse.sc.edu}.
This work was supported in part by the National Security Agency
(NSA) and Advanced Research and Development Activity (ARDA) under Army
Research Office (ARO) contract number DAAD~190210048.} \\
University of South Carolina
\and
Yong Zhang\footnotemark[1]
\\
University of South Carolina}

\date{\today}

\begin{document}

\maketitle

\begin{abstract}
We study two group theoretic problems, {\Gint} and {\DCM}, in the
setting of black-box groups, where {\DCM} generalizes a set of
problems, including {\GM}, {\GF}, and {\CI}.  No polynomial-time
classical algorithms are known for these problems.  We show that for
solvable groups, there exist efficient quantum algorithms for {\Gint}
if one of the underlying solvable groups has a smoothly solvable
commutator subgroup, and for {\DCM} if one of the underlying solvable
groups is smoothly solvable. We also study the decision versions of
{\Stab} and {\OC}, which generalizes {\Gint} and {\DCM}, respectively.
We show that they reduce to {\OS} under certain conditions.  Finally, we
show that {\DCM} and {\DCNM} have zero knowledge proof systems.
\end{abstract}

\section{Introduction}

\begin{figure}
\begin{center}
\input{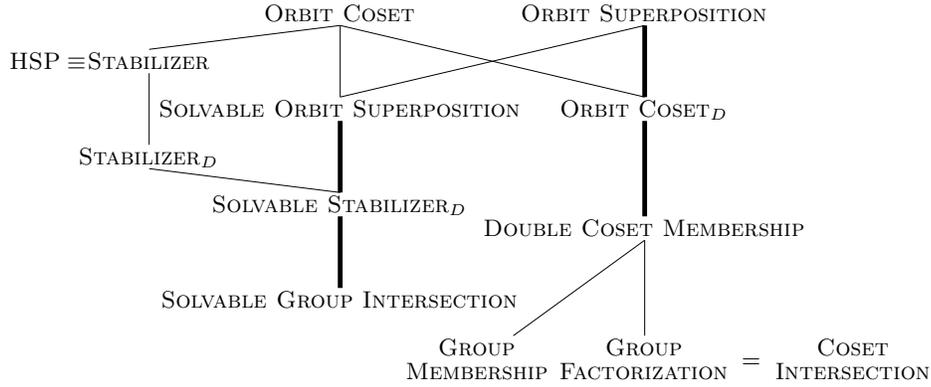}
\caption{Known reducibilities between various group theoretic
problems.  Thick lines represent nontrivial reducibilities shown in
the current work.}\label{fig:group-probs}
\end{center}
\end{figure}

This paper makes progress in finding connections between quantum
computation and computational group theory.  We give results about
quantum algorithms and reductions for group theoretic problems,
concentrating mostly on solvable groups.  These results come in three
sections.  First, we concentrate on two particular group theoretic
problems, {\Gint} and {\DCM}, showing that these problems reduce to
other group problems with known efficient quantum algorithms for many
instances, yielding efficient quantum algorithms for {\Gint} and
{\DCM} on the same types of groups.  Second, we generalize and refine
our results in the first section by introducing decision versions of
the {\Stab} and {\OC} problems (see \cite{friedl03:_hidden}), and
showing that these new problems lie in between {\Gint} and {\DCM} on
the one hand, and the problem {\OS}, defined in
\cite{friedl03:_hidden}, on the other.  Third, we relate our results
on {\DCM} to recent work of Aharonov \& Ta-Shma \cite{aharonov03} by
showing that {\DCM} and its complement have perfect zero knowledge
proofs.  Our results and other known reducibility relationships
between these and other various group theoretic problems are
summarized in Figure~\ref{fig:group-probs}.
A common theme running through all three sections is the surprising
usefulness of producing certain uniform quantum superpositions.

\bigskip

Many problems that have quantum algorithms exponentially faster than
the best known classical algorithms turn out to be special cases of
the {\HS} problem (HSP) for abelian groups, which can be solved using
the Quantum Fourier Transform
\cite{mosca99:_quant_comput_algor,jozsa00:_quant}.  Other interesting
problems, such as {\GI} are special cases of general {\HS}, for which
no efficient quantum algorithm is currently known.  The idea that
underlying algebraic structures may be essential for problems having
exponential quantum speedup has prompted several researchers to study
problems in computational group theory. 
Watrous \cite{watrous01:_quant} first constructed efficient quantum
algorithms for several problems on solvable groups, such as {\OV} and
{\GM}.  Based on an algorithm of Beals and Babai
\cite{beals93:_las_vegas}, Ivanyos, Magniez, and Santha
\cite{ivanyos01:_effic_abelian} obtained efficient quantum algorithms
for {\OV} as well as several other group theoretic problems.
Recently, Friedl et al.\cite{friedl03:_hidden} introduced the problems
{\Stab}, {\OC}, and {\OS}, and showed that these problems can be
solved efficiently on quantum computers if the underlying groups
satisfy certain stronger solvability criteria.

Watrous asked in \cite{watrous01:_quant} whether there are efficient
quantum algorithms for problems such as {\Gint} and {\CI}.
We show that for solvable groups, there are efficient quantum
algorithms for {\Gint} and {\DCM} (which generalizes {\CI} as well as {\GM} and
{\GF}) under certain conditions.  We obtain
these results by showing that these two problems reduce to {\Stab} and
{\OC}, respectively.

One key component in our proof is the
construction of approximately uniform quantum superpositions over
elements of a given solvable group, which is a very useful byproduct
of \cite{watrous01:_quant}.  In classical computational group theory,
the ability to sample group elements uniformly at random is very
useful in designing many classical group algorithms.  We believe that
its quantum analog---uniform quantum superpositions over group
elements---will continue to be useful in designing quantum group
algorithms.
Our results also imply that for \emph{abelian} groups, {\Gint} and
{\DCM} are in the complexity class $\BQP$, which yields a new proof
that they are low for the class $\PP$ \cite{arvind97:_solvab_pp,FR:BQP}.

We observe that in the reduction from {\Gint} (respectively {\DCM}) to
{\Stab} (respectively {\OC}), we don't actually need the full power of
{\Stab} or {\OC}.  This inspires us to study simplified versions of
these two problems.  Here we use {\Stab}$_{D}$ and {\OC}$_{D}$ to
denote the decision versions of these two problems, where we are only
interested in a trivial/non-trivial answer.  We show that the
difficulty of {\Stab}$_{D}$ and {\OC}$_{D}$ may reside in
constructions of certain uniform quantum superpositions, which can be
achieved by the problem {\OS}. In particular, we show that for
solvable groups, {\Stab}$_{D}$ reduces to {\OS}, and for any finite
groups, {\OC}$_{D}$ reduces to {\OS} in bounded-error quantum polynomial time. This
again reinforces our idea that certain uniform quantum superpositions
are key components in quantum group algorithms.

A recent paper by Aharonov and Ta-Shma \cite{aharonov03} shares a
similar point of view.  They studied the problem {\CQS} ($\cqs$),
which basically concerns generating quantum states
corresponding to classical probability distributions.  Furthermore,
they showed interesting connections between $\cqs$ and many different
areas such as Statistical Zero Knowledge ($\SZK$) and adiabatic
evolution.  In particular, they showed that any language in $\SZK$ can
be reduced to a family of instances of $\cqs$. Inspired by this, we
obtain connections between our group theoretic problems and the
complexity class $\SZK$\@. We show that {\DCM} has a zero knowledge
proof system, therefore it is in $\SZK$.  This is an improvement of
Babai's result \cite{babai92:_bound} that {\DCM} is in $\AM \cap
\co{\AM}$. We also give an explicit zero knowledge proof system for
the complement of {\DCM}, namely, {\DCNM}.  While Watrous
\cite{Watrous:BlackBoxGroups} showed that \prob{Group Nonmembership}
is in the complexity class $\QMA$, another implication of our results
is that \prob{Group Nonmembership} has a zero knowledge interactive
proof system.




\section{Preliminaries}
\label{sec:prelims}

Background on general group theory and quantum computation can be
found in the standard textbooks \cite{burnside55,NC:quantumbook}.

\subsection{The Black-Box Group Model}

All of the group theoretic problems discussed in this paper will be
studied in the model of black-box groups.  This model was first
introduced by Babai and Szemer\'{e}di \cite{babai84:matrix} as a
general framework for studying algorithmic problems for finite
groups. It has been extensively studied (see
\cite{watrous01:_quant}). Here we will use descriptions similar to
those in \cite{arvind97:_solvab_pp}.

We fix the alphabet $\Sigma = \{0,1\}$. A \emph{group family} is a
countable sequence ${\cal B}=\{ B_m \}_{m\geq 1}$ of finite groups
$B_m$, such that there exist polynomials $p$ and $q$ satisfying the
following conditions. For each $m\geq 1$, elements of $B_m$ are
encoded as strings (not necessarily unique) in $\Sigma^{p(m)}$. The
group operations (inverse, product and identity testing) of $B_m$ are
performed at unit cost by black-boxes (or group oracles).  The order
of $B_m$ is computable in time bounded by $q(m)$, for each $m$. 
We refer to the groups $B_m$ of a group family and their subgroups
(presented by generator sets) as \emph{black-box groups}.  Common
examples of black-box groups are $\{ S_n \}_{n\geq 1}$ where $S_n$ is
the permutation group on $n$ elements, and $\{GL_n(q)\}_{n\geq 1}$
where $GL_n(q)$ is the group of $n \times n$ invertible matrices over
the finite field $F_q$.  Depending on whether the group elements are
uniquely encoded, we have the \emph{unique encoding model} and 
\emph{non-unique encoding model}, the latter of which enables us to deal with factor
groups \cite{babai84:matrix}.  In the non-unique encoding model an
additional group oracle has to be provided to test if two strings
represent the same group element.  Our results will apply only to the
unique encoding model.  In one of our proofs, however, we will use the
non-unique encoding model to handle factor groups.  For how to
implement group oracles in the form of quantum circuits, please see
\cite{watrous01:_quant}.

\begin{definition}[\cite{arvind97:_solvab_pp}]
Let ${\cal B} = \{ B_m \}_{m\geq 1}$ be a group family. Let $e$ denote
the identity element of each $B_m$.  Let $\tuple{S}$ denote the group
generated by a set $S$ of elements of $B_m$.  Below, $g$ and $h$ denote
elements, and $S_1$ and $S_2$ subsets, of $B_m$.
\begin{eqnarray*}
\mbox{\Gint} & := & \{(0^m, S_1,S_2) \mid \tuple{S_1} \cap \tuple{S_2}
\neq \tuple{e} \}, \\
\mbox{\GM} & := & \{(0^m, S_1, g) \mid g\in \tuple{S_1} \}, \\
\mbox{\GF} & := & \{(0^m, S_1,S_2, g) \mid g \in \tuple{S_1}
\tuple{S_2}  \}, \\
\mbox{\CI} & := & \{(0^m, S_1,S_2,g) \mid \tuple{S_1} g \cap \tuple{S_2}
\neq \emptyset \}, \\
\mbox{\DCM} & := & \{(0^m, S_1,S_2,g,h) \mid g\in \tuple{S_1} h
\tuple{S_2} \}.
\end{eqnarray*}
\end{definition}

It is easily seen that {\DCM} generalizes {\GM}, {\GF}, and
{\CI}. Therefore in this paper we will focus on {\DCM}. All our
results about {\DCM} will also apply to {\GM}, {\GF}, and {\CI}.
(Actually, {\CI} and {\GF} are easily seen to be the same problem.)

\subsection{Solvable Groups}

The \emph{commutator subgroup} $G'$ of a group $G$ is the subgroup
generated by elements $g^{-1}h^{-1}gh$ for all $g,h\in G$.
We define $G^{(n)}$ such that
\begin{eqnarray*}
G^{(0)} & = & G, \\
G^{(n)} & = & (G^{(n-1)})', \mbox{ for } n\geq 1.
\end{eqnarray*}
$G$ is \emph{solvable} if $G^{(n)}$ is the trivial group $\{e\}$ for
some $n$.  We call $G=G^{(0)} \triangleright G^{(1)} \triangleright
\cdots \triangleright G^{(n)}=\{e\}$ the \emph{derived series} of $G$,
of length $n$.  Note that all the factor groups $G^{(i)}/G^{(i+1)}$
are abelian. There is a randomized procedure that computes the derived
series of a given group $G$ \cite{babai95}.

The term \emph{smoothly solvable} is first introduced in
\cite{friedl03:_hidden}.  We say that a family of abelian groups is
\emph{smoothly abelian} if each group in the family can be expressed as
the direct product of a subgroup whose exponent is bounded by a
constant and a subgroup of polylogarithmic size in the order of the
group. A family of solvable groups is \emph{smoothly solvable} if the
length of each derived series is bounded by a constant and the family
of all factor groups $G^{(i)}/G^{(i+1)}$ is smoothly abelian.

In designing efficient quantum algorithms for computing the order of a
solvable group ({\OV}), Watrous \cite{watrous01:_quant} obtained as a
byproduct a method to construct approximately uniform quantum
superpositions over elements of a given solvable group.

\begin{theorem}[\cite{watrous01:_quant}]\label{thm:watrous}
In the model of black-box groups with unique encoding, there is a
quantum algorithm operating as follows (relative to an arbitrary group
oracle).  Given generators $g_1,\ldots,g_m$ such that
$G=\tuple{g_1,\ldots,g_m}$ is solvable, the algorithm outputs the
order of $G$ with probability of error bounded by $\epsilon$ in time
polynomial in $mn+\log (1/\epsilon)$ (where $n$ is the length of the
strings representing the generators). Moreover, the algorithm produces
a quantum state $\rho$ that approximates the state
$\ket{G}=|G|^{-1/2}\sum_{g\in G}\ket{g}$ with accuracy $\epsilon$ (in
the trace norm metric).
\end{theorem}

\subsection{{\Stab}, {\OC} and {\OS}}

A recent paper by Friedl et al.\ \cite{friedl03:_hidden} introduced
several problems which are closely related to {\HS}. In particular,
they introduced {\Stab}, \prob{Hidden Translation}, {\OC}, and {\OS}.
{\Stab} generalizes {\HS}. In fact, the only difference between
{\Stab} and {\HS} is that in the definition of {\Stab} the function
$f$ can be a \emph{quantum function} that maps group elements to
mutually orthogonal
quantum states with unit norm.  {\OC} generalizes {\Stab} and
\prob{Hidden Translation}. {\OS} is a relevant problem, which is also
of independent interest. The superpositions Watrous constructed in
Theorem~\ref{thm:watrous} can be considered as an instance of {\OS}.

In the following we will state the problems and results that will be
used in this paper. We refer interested readers to their paper
\cite{friedl03:_hidden} for detailed information.

Let $G$ be a finite group. Let $\Gamma$ be a set of mutually
orthogonal quantum states. Let $\map{\alpha}{G\times\Gamma}{\Gamma}$
be a group action of $G$ on $\Gamma$, i.e., for every $x\in G$ the
function $\map{\alpha_x}{\ket{\phi}}{\ket{\alpha (x,\ket{\phi})}}$ is
a permutation over $\Gamma$ and the map $h$ from $G$ to the symmetric
group over $\Gamma$ defined by $h(x)=\alpha_x$ is a homomorphism. We
use the notation $\ket{x\cdot \phi}$ instead of $\ket{\alpha
(x,\ket{\phi})}$, when $\alpha$ is clear from the context.  We let
$G(\ket{\phi})$ denote the set $\{\ket{x\cdot\phi}: x\in G\}$, and we
let $G_{\ket{\phi}}$ denote the stabilizer subgroup of $\ket{\phi}$ in
$G$, i.e., $\{x\in G: \ket{x\cdot \phi} = \ket{\phi} \}$.  Given any
positive integer $t$, let $\alpha^t$ denote the group action of $G$ on
$\Gamma^t=\{\ket{\phi}^{\otimes t} : \ket{\phi}\in \Gamma\}$ defined
by $\alpha^t (x,\ket{\phi}^{\otimes t}) = \ket{x\cdot\phi}^{\otimes
t}$. We need $\alpha^t$ because the input superpositions cannot be
cloned in general.

\begin{definition}[\cite{friedl03:_hidden}]
Let $G$ be a finite group and $\Gamma$ be a set of mutually orthogonal
quantum states. Fix the group action
$\map{\alpha}{G\times\Gamma}{\Gamma}$.
\begin{itemize}
\item
Given generators for $G$ and a quantum states $\ket{\phi}\in \Gamma$,
the problem {\Stab} is to find a generating set for the subgroup
$G_{\ket{\phi}}$.
\item
Given generators for $G$ and two quantum states $\ket{\phi_0},
\ket{\phi_1}\in \Gamma$, the problem {\OC} is to either reject the
input if $G(\ket{\phi_0})\cap G(\ket{\phi_1})=\emptyset$ or output a
generating set for $G_{\ket{\phi_1}}$ of size $O(\log |G|)$ and a
$u\in G$ such that $\ket{u\cdot\phi_1}=\ket{\phi_0}$.
\item
Given generators for $G$ and a quantum state $\ket{\phi}\in \Gamma$,
the problem {\OS} is to construct the uniform superposition
\[ \ket{G\cdot\phi} = \frac{1}{\sqrt{|G(\ket{\phi})|}}\sum_{\ket{\phi'}\in
G(\ket{\phi})} \ket{\phi'}. \]
\end{itemize}
\end{definition}

{\OC} and {\Stab} can be solved in quantum polynomial time 
under certain stronger solvability criteria.
\begin{theorem}[\cite{friedl03:_hidden}]\label{thm:orbit}
Let $G$ be a smoothly solvable group and let $\alpha$ be a group
action of $G$. When $t=(\log^{\Omega (1)} |G|)\log (1/\epsilon)$,
{\OC} can be solved in $G$ for $\alpha^t$ in quantum time poly$(\log
|G|)\log(1/\epsilon)$ with error $\epsilon$.
\end{theorem}

\begin{theorem}[\cite{friedl03:_hidden}]\label{thm:stab}
Let $G$ be a finite solvable group having a smoothly solvable
commutator subgroup and let $\alpha$ be a group action of $G$. When
$t=(\log^{\Omega (1)} |G|)\log (1/\epsilon)$, {\Stab} can solved in
$G$ for $\alpha^t$ in quantum time poly$(\log |G|)\log(1/\epsilon)$
with error $\epsilon$.
\end{theorem}

Another interesting result in \cite{friedl03:_hidden} is that {\OS}
reduces to {\OC} for solvable groups in quantum polynomial time. 
It is not clear if there is a reduction in the reverse direction.

\subsection{Zero Knowledge Proof Systems}

We use standard notions of interactive proof systems and
zero knowledge interactive proof systems.  Information about
zero knowledge systems can be found in a variety of places, including
Vadhan's Ph.D. thesis \cite{vadhan99}, and Goldreich, Micali, \&
Wigderson \cite{goldreich91:_proof_np}.

$\SZK$ is the class of languages that have statistical zero knowledge
proofs.  It is known that $\BPP \subseteq \SZK \subseteq \AM \cap
\co{\AM}$ and that $\SZK$ is closed under complement. $\SZK$ does not
contain any $\NP$-complete language unless the polynomial hierarchy
collapses \cite{vadhan99}.

\subsection{A Note on Quantum Reductions}

In Sections~\ref{sec:q-algos} and \ref{sec:decision-probs} we describe
quantum reductions to various problems.  Quantum algorithms for these
problems often require several identical copies of a quantum state or
unitary gate to work to a desired accuracy.  Therefore, we will
implicitly assume that our reductions may be repeated $t$ times, where
$t$ is some appropriate parameter polynomial in the input size and the
logarithm of the desired error bound.

\section{Quantum algorithms}
\label{sec:q-algos}

In this section we report progress on finding quantum algorithms for {\Gint}, 
and {\DCM}.

\begin{theorem}\label{thm:gi}
{\Gint} reduces to {\Stab} in bounded-error quantum polynomial time if
one of the underlying groups is solvable.
\end{theorem}

\begin{proof}
Given an input $(0^m,S_1,S_2)$ for {\Gint}, without loss of generality, 
suppose that $G=\tuple{S_1}$ is an arbitrary finite group and $H=\tuple{S_2}$ is solvable.
By Theorem~\ref{thm:watrous} we can construct an approximately uniform
superposition $\ket{H}=|H|^{-1/2}\sum_{h\in H}\ket{h}$. For any $g\in
G$, let $\ket{gH}$ denote the uniform superposition over left coset
$gH$, i.e., $\ket{gH}=|H|^{-1/2}\sum_{h\in gH}\ket{h}$.  Let $\Gamma =
\{ \ket{gH} |g \in G\}$. Note that the quantum states in $\Gamma$ are
(approximately) pairwise orthogonal.  Define the group action
$\map{\alpha}{G \times \Gamma}{\Gamma}$ to be that for every $g\in G$
and every $\ket{\phi} \in \Gamma$, $\alpha (g, \ket{\phi}) =
\ket{g\phi}$.  Then the intersection of $G$ and $H$ is exactly the
subgroup of $G$ that stabilizes the quantum state $\ket{H}$.
\end{proof}

\begin{corollary}\label{cor:gi}
{\Gint} over solvable groups can be solved within error $\epsilon$ by
a quantum algorithm that runs in time polynomial in $m +
\log(1/\epsilon)$, where $m$ is the size of the input, provided one of
the underlying solvable groups has a smoothly solvable commutator
subgroup.
\end{corollary}

\begin{proof}
Follows directly from Theorems~\ref{thm:gi} and \ref{thm:stab}.
\end{proof}

It is not clear if similar reduction to {\Stab} exists for {\DCM}.
However, with the help of certain uniform superpositions, {\DCM} can be nicely 
put into the framework of {\OC}.

\begin{theorem}\label{thm:dcm}
{\DCM} over solvable groups reduces to {\OC} in bounded-error
quantum polynomial time.
\end{theorem}

\begin{proof}
Given input for {\DCM} $S_1$, $S_2$, $g$ and $h$, where
$G=\tuple{S_1}$ and $H=\tuple{S_2}$ are solvable groups, first we
check if $g$ is an element of $G$ or $H$. This can be done using the
quantum algorithm for {\GM} in \cite{watrous01:_quant}.  For example,
to check if $g$ is an element of $G$, the algorithm will check if the
group $\tuple{S_1,g}$ is still solvable, and in the case that it is
solvable compute the order of $\tuple{S_1,g}$ and check if it is equal
to the order of $G$. If $g$ is an element of $G$ or $H$, quit and
output ``yes.''

In the case that $g$ is not an element of $G$ or $H$, we construct the
input for {\OC} as follows. Let $\Gamma=\{ \ket{xH} |x\in \tuple{S_1,
S_2,g, h}\}$. Define group action
$\map{\alpha}{G\times\Gamma}{\Gamma}$ to be $\alpha
(x,\ket{\phi})=\ket{x\phi}$ for any $x\in G$ and $\ket{\phi}\in
\Gamma$.  Let two input quantum states $\ket{\phi_0}$ and
$\ket{\phi_1}$ be $\ket{gH}$ and $\ket{hH}$, which can be constructed using
Theorem~\ref{thm:watrous}.  It is not hard to check that there exists
an $u\in G$ such that $\ket{u\cdot \phi_1}=\ket{\phi_0}$ if and only
if $g \in GhH$.
\end{proof}

\begin{corollary}\label{cor:dcm}
{\DCM} over solvable groups can be solved within error $\epsilon$ 
by a quantum algorithm that runs in time polynomial in 
$m + \log(1/\epsilon)$, where $m$ is the size of the input,
provided one of the underlying groups is smoothly solvable.
\end{corollary}

\begin{proof}
Given input for {\DCM} $S_1$, $S_2$, $g$ and $h$, suppose that 
$G=\tuple{S_1}$ is smoothly solvable and $H=\tuple{S_2}$ is solvable.
Let $S_1, \ket{gH}, \ket{hH}$ be the input for {\OC}, the result follows 
from Theorem~\ref{thm:orbit}. If instead $H$
is the one which is smoothly solvable, then we modify the input
by swapping $S_1$ and $S_2$ and using $g^{-1}, h^{-1}$ to replace $g, h$.
Note that this modification will not change the final answer.
\end{proof}

\section{The decision versions of {\Stab} and {\OC}}
\label{sec:decision-probs}

An interesting observation is that to solve our group theoretic
problems, we don't actually need the full power of {\Stab} and
{\OC}. For example, for the problem {\Gint}, we care about whether the
intersection of the two input groups is trivial or non-trivial. We
don't ask for a generating set in the case of a non-trivial
intersection.  This inspires us to define and study the decision
versions of {\Stab} and {\OC}.  denoted as {\Stab}$_{D}$ and
{\OC}$_{D}$, respectively. 

\begin{definition}
Let $G$ be a finite group and $\Gamma$ be a set of pairwise orthogonal
quantum states. Fix the group action
$\map{\alpha}{G\times\Gamma}{\Gamma}$.
\begin{itemize}
\item
Given generators for $G$ and a quantum state
$\ket{\phi}\in \Gamma$, the problem {\Stab}$_{D}$ is to check if the
subgroup $G_{\ket{\phi}}$ is the trivial subgroup $\{ e\}$.
\item 
Given generators for $G$ and two quantum states
$\ket{\phi_0}, \ket{\phi_1}\in \Gamma$, the problem {\OC}$_{D}$ is to
either reject the input if $G(\ket{\phi_0})\cap
G(\ket{\phi_1})=\emptyset$ or accept the input if $G(\ket{\phi_0}) =
G(\ket{\phi_1})$.
\end{itemize}
\end{definition} 
 
It is clear that the reductions in Theorem~\ref{thm:gi} and
Theorem~\ref{thm:dcm} still work if we replace {\Stab} (respectively
{\OC}) with {\Stab}$_{D}$ (respectively {\OC}$_{D}$).  We remark that
although {\OC} generalizes {\Stab}, {\OC}${_D}$ does not seem to
generalize {\Stab}${_D}$.  Next we show that the ability of
constructing certain quantum superpositions will help us to attack
these two problems. The problem {\OS} provides a way to construct
quantum superpositions. In fact, Watrous' result in
Theorem~\ref{thm:watrous} solves a special case of {\OS}, where the
group $G$ acts on the quantum state of the identity element.

We will use the following result from \cite{ivanyos01:_effic_abelian}:  

\begin{theorem}[\cite{ivanyos01:_effic_abelian}]\label{thm:ims}
Assume that $G$ is a black-box group given by generators with not necessarily 
unique encoding. Suppose that $N$ is a normal
subgroup given as a hidden subgroup of $G$ via the function $f$. 
Then the order of the factor group $G/N$ can be computed
by quantum algorithms in time polynomial in 
$n+\nu(G/N)$, where $n$ is the input size and the parameter $\nu(G)$ is defined
in \cite{beals93:_las_vegas} and equals one for any solvable group $G$.
\end{theorem}

Please note that we can apply Theorem~\ref{thm:ims} to factor groups since it uses the 
non-unique encoding black-box groups model.

\begin{theorem}\label{thm:stabilizer}
Over solvable groups, {\Stab}$_{D}$ reduces to {\OS} in bounded-error 
quantum polynomial time.
\end{theorem}

\begin{proof}
Let the solvable group $G$ and quantum state $\ket{\phi}$ be the input
of {\Stab}$_{D}$.  We can find in classical polynomial time generators
for each element in the derived series of $G$ \cite{babai95}, namely,
$\{e\}=G_1\triangleleft \cdots \triangleleft G_n=G$.  For $1\leq i\leq
n$ let $S_i = (G_i)_{\ket{\phi}}$, the stabilizer of $\ket{\phi}$ in
$G_i$.  By Theorem~\ref{thm:watrous} we can compute the orders of $G_1,
\ldots, G_n$ and thus the order of $G_{i+1}/G_i$ for any $1\leq i <
n$.  We will proceed in steps.  Suppose that before step $i+1$, we
know that $S_i =\{e\}$.  We want to find out if
$S_{i+1} =\{e\}$ in the $(i+1)$st step.  Since $G_i
\triangleleft G_{i+1}$, by the Second Isomorphism Theorem,
$G_iS_{i+1}/ G_i \isoto S_{i+1}$.
Consider the factor group $G_{i+1}/ G_i$, we will define a function
$f$ such that $f$ is constant on $G_iS_{i+1}/ G_i$ and
distinct on left cosets of $G_iS_{i+1}/ G_i$ in
$G_{i+1}/ G_i$.  Then by Theorem~\ref{thm:ims} we can compute the
order of the factor group $G_{i+1}/ G_i$ over
$G_iS_{i+1}/ G_i$.  The group oracle needed
in the non-unique encoding model to test if two strings $s_1$ and
$s_2$ represent the same group elements can be implemented using the
quantum algorithm for {\GM}, namely, testing if $s_1^{-1}s_2$ is a
member of $G_i$. The order of this group is equal to the order of
$G_{i+1}/ G_i$ if and only if $S_{i+1}$ is trivial.

Here is how we define the function $f$.  Using $G_i$ and $\ket{\phi}$
as the input for {\OS}, we can construct the uniform superposition
$\ket{G_i\cdot\phi}$.  Let $\Gamma$ be the set $\{\ket{gG_i\cdot\phi}
| g\in G_{i+1} \}$.  We define $\map{f}{G_{i+1}/G_i}{\Gamma}$ be such
that $f(gG_i)=\ket{gG_i\cdot\phi}$.  What is left is to verify that
$f$ hides the subgroup $G_iS_{i+1} / G_i$ in the group $G_{i+1}/ G_i$.
For any $g\in G_iS_{i+1}$, it is straightforward to see that
$\ket{gG_i\cdot\phi} =\ket{G_i\cdot\phi}$. If $g_1$ and $g_2$ are in
the same left coset of $G_iS_{i+1}$, then $g_1 = g_2
g$ for some $g\in G_iS_{i+1}$ and thus
$\ket{g_1G_i\cdot\phi} =\ket{g_2G_i\cdot\phi}$. If $g_1$ and $g_2$ are not in
the same left coset of $G_iS_{i+1}$, we will show
that $\ket{g_1G_i\phi}$ and $\ket{g_2G_i\phi}$ are orthogonal quantum
states. Suppose there exists $x_1, x_2\in G_i$ such that
$\ket{g_1x_1\cdot\phi}=\ket{g_2x_2\cdot\phi}$,  then $x_1^{-1}g_1^{-1}g_2x_2 \in
S_{i+1}$. But $x_1^{-1}g_1^{-1}g_2x_2 =
x_1^{-1}x_2'g_1^{-1}g_2$ for some $x_2' \in G_i$. Thus
$g_1^{-1}g_2 \in G_iS_{i+1}$.  This contradicts the
assumption that $g_1$ and $g_2$ are not in the same coset of
$G_iS_{i+1}$.

We need to repeat the above procedure at most $\Theta(\log |G|)$
times. For each step the running time is polynomial in $\log |G|+\log(1/\epsilon)$,
for error bound $\epsilon$. So
the total running time is still polynomial in the input size.
\end{proof}


\begin{corollary}\label{cor:gi_to_os}
Over solvable groups, {\Gint} reduces to {\OS} in
bounded-error quantum polynomial time.
\end{corollary}

We can also reduce {\OC}${_D}$ to {\OS} in quantum polynomial time.  
In this reduction, we
don't require the underlying groups to be solvable. The proof uses 
similar techniques that Watrous \cite{Watrous:BlackBoxGroups} and
Buhrman et al.\ \cite{buhrman:_quantum_fingerprinting}
used to differentiate two quantum states.

\begin{theorem}\label{thm:orbitcoset}
{\OC}$_{D}$ reduces to {\OS} in bounded-error quantum polynomial time.
\end{theorem}

\begin{proof}
Let the finite group $G$ and two quantum states $\ket{\phi_1}$,
$\ket{\phi_2}$ be the inputs of {\OC}$_{D}$. Notice that the orbit
coset of $\ket{\phi_1}$ and $\ket{\phi_2}$ are either identical or
disjoint, which implies the two quantum states $\ket{G\cdot\phi_1}$ and
$\ket{G\cdot\phi_2}$ are either identical or orthogonal.  We may then tell
which is the case using a version of the swap test of Buhrman et al.\
\cite{buhrman:_quantum_fingerprinting}.

%
\end{proof}  

\begin{corollary}
{\DCM} reduces to {\OS} in bounded-error quantum polynomial time.
\end{corollary}

\section{Statistical Zero Knowledge}
\label{sec:zk}

A recent paper by Aharonov and Ta-Shma \cite{aharonov03}
proposed a new way to generate certain quantum states using Adiabatic
quantum methods. In particular, they introduced the problem
\prob{Circuit Quantum Sampling} (CQS) and its connection to the
complexity class Statistical Zero Knowledge ($\SZK$). Informally
speaking, CQS is to generate quantum states corresponding to classical
probability distributions obtained from some classical circuits.
Although CQS and {\OS} are different problems, they bear a certain
level of resemblance.  Both problems are concerned about generation of
non-trivial quantum states.  In their paper they showed that any
language in $\SZK$ can be reduced to a family of instances of CQS.
Based on Theorem~\ref{thm:stabilizer} and
Theorem~\ref{thm:orbitcoset}, We would like to ask if there are
connections between $\SZK$ and our group theoretic problems.  As a
first step, we show that {\DCM} has a perfect zero knowledge proof
system, and thus is in $\SZK$.  This is an improvement of Babai's
result \cite{babai92:_bound} that {\DCM} is in $\AM \cap
\co{\AM}$. Our proof shares the same flavor with Goldreich, Micali and
Wigderson's proof that {\sc Graph Isomorphism} is in $\SZK$
\cite{goldreich91:_proof_np}.  The intuitive idea is to break the
process into two parts, where the verification of each individual part
does not reveal any information about the claim.

The following theorem due to Babai \cite{babai91:_local} will be used
in our proof.  Let $G$ be a finite group.  Let $g_1,\ldots,g_k \in G$
be a sequence of group elements. A {\em subproduct} of this sequence
is an element of the form $g_1^{e_1}\ldots g_k^{e_k}$, where $e_i \in
\{ 0,1 \}$. We call a sequence $h_1,\ldots,h_k \in G$ {\em a sequence
of $\epsilon$-uniform Erd\H{o}s-R\'{e}nyi generators} if every element
of $G$ is represented in $(2^k/|G|)(1+\epsilon)$ ways as a subproduct
of the $h_i$.
\begin{theorem}[\cite{babai91:_local}]\label{Babai:rand}
Let $c, C>0$ be given constants, and let $\epsilon=N^{-c}$ where $N$
is a given upper bound on the order of the group $G$. There is a Monte
Carlo algorithm which, given any set of generators of $G$, constructs
a sequence of $O(\log N)$ $\epsilon$-uniform Erd\H{o}s-R\'{e}nyi
generators at a cost of $O((\log N)^5)$ group operations. The
probability that the algorithm fails is $\leq N^{-C}$. If the
algorithm succeeds, it permits the construction of $\epsilon$-uniform
distributed random elements of $G$ at a cost of $O(\log N)$ group
operations per random element.
\end{theorem}

Basically what Theorem~\ref{Babai:rand} says is that we can randomly
sample elements from $G$ and verify the membership of the random
sample efficiently. Given a group $G$ and a sequence of $O(\log N)$
$\epsilon$-uniform Erd\H{o}s-R\'{e}nyi generators $h_1,\ldots,h_k$ for
$G$, we say that $e_1 \ldots e_k$ where $e_i \in \{ 0,1 \}$ is a {\em
witness} of $g\in G$ if $g=h_1^{e_1}\ldots h_k^{e_k}$.

\begin{theorem}\label{thm:dcmszk}
{\DCM} has a perfect zero knowledge proof system.
\end{theorem}
\begin{proof}[sketch]
Given groups $G$, $H$ and elements $g$, $h$, the prover wants to
convince the verifier that $g=xhy$ for some $x\in G$ and $y\in H$. Fix
a sufficiently small $\epsilon >0$. The protocol is as follows.
\begin{description}
\item[(V0)] The verifier computes $\epsilon$-uniform
Erd\H{o}s-R\'{e}nyi generators $g_1,\ldots,g_m$ and $h_1,\ldots,h_n$
for $G$ and $H$. The verifier sends the generators to the prover.
\item[(P1)] The provers select $x$ and $y$, which are random elements
from $G$ and $H$. The prover sends $z=xgy$ to the verifier.
\item[(V1)] The verifier chooses at random $\alpha \in_{R} \{ 0,1 \}$,
and sends $\alpha$ to the prover.
\item[(P2)] If $\alpha = 0$, then the prover sends $x$ and $y$ to the
verifier, together with witnesses that $x\in G$ and $y\in H$. If
$\alpha = 1$, then the prover sends over $x'$ and $y'$, together with
witnesses that $x'\in G$ and $y'\in H$.
\item[(V2)] If $\alpha = 0$, then the verifier verifies that $x$ and
$y$ are indeed elements of $G$ and $H$ and $z=xgy$. If $\alpha =1$,
then the verifier verifies that $x'$ and $y'$ are indeed elements of
$G$ and $H$ and $z=x'hy'$. The verifier stops and rejects if any of
the verifications fails. Otherwise, he repeats steps from (P1) to
(V2).
\end{description}
If the verifier has completed $m$ iterations of the above steps, then
he accepts.

It is not hard to verify that this is a perfect zero knowledge
proof system. We omit the formal proof due to lack of space.
\end{proof}

Since $\SZK$ is closed under complement, the complement of {\DCM},
{\DCNM}, is also in $\SZK$. In fact, by adapting proofs in
\cite{goldreich91:_proof_np}, we can give explicitly a perfect zero
knowledge proof system for {\DCNM}.

\begin{theorem}\label{thm:DCNM}
{\DCNM} has a perfect zero knowledge proof system.
\end{theorem}

\begin{proof}[sketch]
A simple interactive proof system for {\DCNM} is as follows. Given
$G$, $H$ and $g$, $h$ as inputs, the prover wants to convince the
verifier that $g$ is not in the double coset $GhH$. The verifier will
generate random elements $x \in G$ and $y\in H$, and then flip a
random coin and send either $xgy$ or $xhy$ to the prover. The prover
has to tell correctly which one the verifier sends. After several
rounds, the verifier is convinced. This protocol is not zero knowledge
since a cheating verifier can use the protocol to gain knowledge such
as whether an element $z$ is in the double coset $GgH$. The way to fix
this flaw is to let the verifier first ``prove'' to the prover that he
knows the answer of his own question.

For the sake of simplicity, let $n$ denote the input size.
Given groups $G$, $H$ and elements $g$, $h$, the prover wants to
convince the verifier that $g$ is not in the double coset $GhH$.
Before the protocol starts, the verifier will compute
$\epsilon$-uniform Erd\H{o}s-R\'{e}nyi generators $g_1,\ldots,g_m$ and
$h_1,\ldots,h_n$ for $G$ and $H$ for a sufficiently small $\epsilon$,
and send them to the prover.

The following protocol will be executed $m$ times, each time using
independent random coin tosses.
\begin{description}
\item[(V1)] The verifier computes random elements $x\in G$ and $y\in
H$ using the Erd\H{o}s-R\'{e}nyi generators, and chooses at random
$\alpha \in_{R} \{0,1\}$. If $\alpha=0$, he computes $z=xgy$. If
$\alpha=1$, he computes $z=xhy$. The element $z$ will be called the
{\em question}. In addition to $z$, the verifier constructs $n^2$
pairs of group elements such that each pair consists of one random
element of $GgH$ and one random element of $GhH$.  The two elements in
each pair are placed at random order. These pairs will be used by the
prover to test whether the verifier is cheating. In specific, for each
$1\leq i \leq n^2$, the verifier constructs the $i$'th pair $(T_{i,0},
T_{i,1})$ as follows. He computes random elements $x_{i,0}, x_{i,1}\in
G$ and $y_{i,0}, y_{i,1}\in H$, and chooses at random a bit
$\gamma_i\in_R \{0,1\}$. Then he computes
$T_{i,\gamma_i}=x_{i,\gamma_i}gy_{i,\gamma_i}$ and
$T_{i,1-\gamma_i}=x_{i,1-\gamma_i}gy_{i,1-\gamma_i}$. The verifier
sends $z$ and the sequence of pairs $(T_{1,0}, T_{1,1}), \ldots,
(T_{n^2,0}, T_{n^2, 1})$ to the prover.

\item[(P1)] The prover chooses at random a subset $I\subseteq \{1,\ldots,n^2\}$ (uniformly among
all $2^{n^2}$ subsets) and sends $I$ to the verifier.

\item[(V2)] If $I$ is not a subset of $\{1,\ldots,n^2\}$, then the verifier halts and rejects.
Otherwise, the verifier replies with $\{(\gamma_i, x_{i,0}, x_{i,1}, y_{i,0}, y_{i,1}): i\in I\}$
and $\{(\alpha_i \in \{0,1\}, a_i \in G, b_i\in H) \mbox{ such that } z=a_iT_{i,\alpha_i}b_i : 
i\notin I\}$. Intuitively, for $i\in I$ the verifier shows that the $i$'th pair is properly 
constructed by giving explicitly $(\gamma_i, x_{i,0}, x_{i,1}, y_{i,0}, y_{i,1})$; for 
$i\notin I\}$ the verifier shows that $z$ is also properly constructed by showing that 
$z$ is in the same double coset with one of the elements in the $i$'th pair. 
$(\alpha_i, a_i, b_i)$ can be easily computed by the verifier, i.e., $\alpha_i = (\alpha + \gamma_i)
\mod 2$, $a_i=xx_{i,\alpha_i}^{-1}$, and $b_i=y_{i,\alpha_i}^{-1}y$.

\item[(P2)] For every $i\in I$, the prover checks whether $x_{i,0}, x_{i,1}$ (respectively 
$y_{i,0}, y_{i,1})$) are indeed elements of $G$ (respectively $H$), and whether 
$T_{i,\gamma_i}=x_{i,\gamma_i}gy_{i,\gamma_i}$ and
$T_{i,1-\gamma_i}=x_{i,1-\gamma_i}gy_{i,1-\gamma_i}$ hold.
For every $i\notin I$, the prover checks whether $a_i$ (respectively $b_i$) is indeed an element
of $G$ (respectively $H$), and whether $z=a_iT_{i,\alpha_i}b_i$ holds. If any of these conditions
does not hold, the prover stops. Otherwise, the prover answers with $\beta \in \{0,1\}$.

\item[(V3)] The verifier checks whether $\alpha = \beta$. If the condition is violated, the verifier
stops and rejects; otherwise, he continues.
\end{description}

After $m$ rounds of successful iterations, the verifier accepts.

This is still an interactive proof system for {\DCNM}. If $g$ is not
in the double coset $GhH$, then $GgH$ and $GhH$ are disjoint sets and
the prover will always succeed in convincing the verifier. If, on the
other hand, $g$ is in the double coset $GhH$, then $GgH$ and $GhH$ are
the same set and with probability at least a half the prover will fail
to fool the verifier.

To prove that this protocol is zero knowledge, the simulator has to
produce the same probability distribution without interacting with the
prover. What the simulator does is to extract from the verifier the
knowledge he has about his question. We omit the formal proof here. We
note that the formal proof is similar in principle to the proof that
\prob{Graph Nonisomorphism} has a zero knowledge proof system
\cite{goldreich91:_proof_np}, based on which and the above protocol
interested readers are able to construct the formal proof.
\end{proof}

Although {\Gint} is also known to be in $\AM \cap \co{\AM}$
\cite{babai92:_bound}, it is not clear whether {\Gint} has a zero
knowledge proof system. This seems to be consistent with the fact that
we have not found a reduction from {\Gint} to {\OS} over arbitrary finite
groups (Corollary~\ref{cor:gi_to_os}).

\paragraph{Acknowledgments.} We would like to thank George Mcnulty,
Fr\'{e}d\'{e}ric Magniez, John Watrous, Variyam Vinodchandran, Derek
Robinson, Scott Aaronson, and Dorit Aharonov for many useful
discussions.

\bibliography{groupintersection.bbl}

\end{document}